\documentclass[prl,twocolumn,amsmath,superscriptaddress,amssymb]{revtex4-2}

\usepackage{graphicx}
\usepackage{dcolumn}
\usepackage{bm}
\usepackage{amssymb}
\usepackage{amsmath}
\usepackage{wasysym}
\usepackage{color}
\usepackage{times}

\usepackage{placeins}
\usepackage{epstopdf}

\usepackage[separate-uncertainty=true, multi-part-units=single]{siunitx}

\usepackage{ulem}
\usepackage{nicefrac}
\usepackage[dvipsnames]{xcolor}

\usepackage[colorlinks]{hyperref} % add hypertext capabilities
\hypersetup{breaklinks=true, citecolor=blue, linkcolor=blue, menucolor=blue, urlcolor=blue}

\usepackage{comment}
\usepackage{textcase}

\usepackage{csquotes}

\usepackage[capitalise]{cleveref}

\makeatletter
\renewcommand*\env@matrix[1][c]{\hskip -\arraycolsep
  \let\@ifnextchar\new@ifnextchar
  \array{*\c@MaxMatrixCols #1}}
\makeatother
\begin{document}

\title{Moiré-Tailored Dirac States with Quasi-One-Dimensional Confined Electrons}

\title{Observation of Moiré-Tailored One-Dimensional Dirac Quasiparticles}

\title{Moiré-Tailored Creation and Control of One-Dimensional Dirac Quasiparticles}

\title{Observation of Moiré-Tailored Dirac Quasiparticles with Topological Protection}

\title{Observation and Control of Moiré-Tailored Topological Dirac States}

%\title{Soliton Domain Boundary States and Dirac Quasiparticles Interwined by a Moiré Lattice}

%\title{Observation and Epitaxial Control of Moiré-Soliton Quasiparticle Excitations}

%\title{On the Local Mechanisms Underlying the Formation of Band Gaps in Moiré-Type Superlattices}

\author{R. Ganser}
\thanks{These authors contributed equally to the present work.}

\author{M. P. T. Masilamani}
\thanks{These authors contributed equally to the present work.}

\author{B. Geldiyev}
\affiliation{Experimentelle Physik VII and W\"{u}rzburg-Dresden Cluster of Excellence ctd.qmat, Universit\"{a}t W\"{u}rzburg, Am Hubland, D-97074 W\"{u}rzburg, Germany}

\author{M. M. Hirschmann}
\affiliation{RIKEN Center for Emergent Matter Science (CEMS), Wako, Saitama 351-0198, Japan}

\author{A. Consiglio}
\affiliation{CNR-IOM Istituto Officina dei Materiali, I-34139 Trieste, Italy}

\author{J. Schusser}
\affiliation{New Technologies-Research Center, University of West Bohemia, 30614, Pilsen, Czech Republic}

\author{D. Di Sante}
\affiliation{Department of Physics and Astronomy, Alma Mater Studiorum, University of Bologna, 40127 Bologna, Italy}

\author{M. \"{U}nzelmann}
\email{maximilian.uenzelmann@uni-wuerzburg.de}
\affiliation{Experimentelle Physik VII and W\"{u}rzburg-Dresden Cluster of Excellence ctd.qmat, Universit\"{a}t W\"{u}rzburg, Am Hubland, D-97074 W\"{u}rzburg, Germany}

%\author{G. Sangiovanni}
%\affiliation{Institut f\"{u}r Theoretische Physik und Astrophysik and W\"{u}rzburg-Dresden Cluster of Excellence ctd.qmat, Universit\"{a}t W\"{u}rzburg, Am Hubland, D-97074 W\"{u}rzburg, Germany}

\author{F. Reinert}
\affiliation{Experimentelle Physik VII and W\"{u}rzburg-Dresden Cluster of Excellence ctd.qmat, Universit\"{a}t W\"{u}rzburg, Am Hubland, D-97074 W\"{u}rzburg, Germany}

\date{\today}

\begin{abstract}
Moiré heterostructures provide a powerful framework for tailoring electronic band structures via controlled long-range periodic superlattice potentials.
%Mostly relying on mechanical stacking, moiré systems --- such as twisted van der Waals layers --- exhibit correlation effects that emerge from almost completely flat bands in the folded moiré band structure.
%Beyond that, replica bands can form emergent Dirac states, which have recently attracted interest.
Beyond widely studied moiré-tailored flat bands, folded band structures can host emergent Dirac states, which have recently attracted considerable interest.
Direct momentum-resolved observation of gapless moiré-Dirac quasiparticles, however, is challenging and has so far remained elusive.
%likewise absent in the parent materials.
%However, this perspective has been barely explored in previous studies.
By performing angle-resolved photoemission spectroscopy measurements on an epitaxial
%one-dimensional
surface-moiré structure, we here provide direct spectroscopic evidence of moiré-dressed Dirac states with topological character. 
%In these, the electrons are strongly trapped by a line-moiré superlattice, identified by the presence of band gaps. With significantly lowered kinetic energy, they in turn propagate as massless Dirac quasiparticles.
Driven by the one-dimensional superlattice potential, electrons propagate anisotropically with a weak but massless Dirac dispersion along the confinement direction. 
The observed band crossings belong to topological nodal lines pinned to the mini-Brillouin zone boundaries. As such, they are enforced and robustly protected by the non-symmorphic symmetry of the superlattice. Finally, we demonstrate that the topological excitations can be almost continuously controlled by tuning the moiré lattice periodicity, directly unveiling 
%means of epitaxy as well as sample temperature. Overall, our work suggests moiré lattices 
moiré heterostructures as a promising platform for creating and controlling topological moiré-Dirac states.
%with topological protection. 
%that can possibly be intertwined with electron-correlation effects.
\end{abstract}

\maketitle

Dirac quasiparticle excitations in crystalline solids belong to one of the key paradigms in quantum matter research \cite{CastroNeto2009, Wehling2014}. They have been observed in various settings, starting from graphene \cite{Novoselov2005, Ohta2006, Bostwick2007} over topological insulators \cite{Review_TI_1, Review_TI_2} to Dirac- and Weyl semimetals \cite{Young2012, Weng2015, Burkov2016, Armitage2018, Xie2021, Uenzelmann2021}. Moreover, non-symmorphic crystal symmetries can generate topological Dirac nodal lines (DNL) with particularly robust symmetry protection \cite{Young2015, Schoop2016, Wilde2021, Hirschmann2021}. DNL have recently been shown to exhibit orbital quantum vortices \cite{Figgemeier2025}, dissipationless transport signatures \cite{Veyrat2025}, and are a central subject of recent theory developments \cite{Rhim2026}. 
In all the cases given, electronic Dirac fermions are commonly formed on the atomic crystal lattice of \textit{natural} quantum materials, where the short-range atomic lattice potential and associated symmetries shape the electronic band structure. For this reason, the controlled manipulation of Dirac states is a complex challenge and typically requires material chemistry approaches.

Meanwhile, artificial moir{\'{e}} heterostructures have emerged as a promising platform for creating controllable electronic structures \cite{Mak2022, Mele2010, Suarez2010, Bistritzer2011, Wu2018, Wu2019, Balents2020, Cao2018_1, Cao2018_2, Li2021, Wang2022, Utama2021, Lisi2021, Gatti2023, Zhang2025, Jiang2025, Yang2025, Ma2025, Herbut2006, Biedermann2025, Biedermann2026}.
In particular, moiré engineering enables the tailoring of correlated states that are absent in the parent materials. The emergence of a long-range superlattice potential gives rise to band gaps in the folded moir{\'{e}} band structure (so-called minibands), thereby quenching kinetic energy and increasing electron-electron interactions.
Emergent flat bands have been observed in advanced angle-resolved photoemission spectroscopy (ARPES) experiments \cite{Utama2021, Lisi2021, Gatti2023, Zhang2025, Jiang2025} and demonstrated to trigger correlation-driven phenomena, like unconventional superconductivity \cite{Cao2018_1}, correlated insulators \cite{Cao2018_2}, the quantum anomalous Hall effect \cite{Li2021}, and Luttinger liquids \cite{Wang2022}.

Beyond flat bands, moir{\'{e}} superlattices have been proposed to form Dirac quasiparticles in the replica band structure. These \textit{slow Dirac fermions} can, in turn, be accompanied by strong electron-electron interaction effects, also referred to as relativistic Mott transition \cite{Yang2025, Ma2025, Herbut2006, Biedermann2025, Biedermann2026}. Using transport, this has recently been demonstrated in artificial honeycomb lattices that emerge in transition metal dichalcogenide moir{\'{e}} heterostructures \cite{Yang2025, Ma2025}. Clear momentum-resolved evidence for the associated massless Dirac quasiparticles, however, is challenging and has so far remained elusive.
Particular challenges are: First, access to mini-bands in van der Waals heterostructures usually requires ARPES experiments with state-of-the-art (sub)-micrometer spatial resolution. Despite fantastic recent methodological developments on this (see e.g. Refs.~\cite{Utama2021, Lisi2021, Gatti2023, Zhang2025, Jiang2025}), these experiments often pay with a reduced energy resolution reaching the order of the entire expected mini-bandwidth, which would then lead to an obstruction of the Dirac crossings within the broadened linewidth. 
Second, Dirac states, which are not robustly protected and thus unstable to small perturbations or disorder, may easily gap out \cite{Wang2016}, repressing their massless character.
Third, and probably most crucially, the detection of gapless mini-band Dirac crossings in ARPES is very delicate, because next to actual initial-state replica bands, photo-excited quasi-free electrons can also scatter at surface superstructures. This likewise gives rise to generically gapless backfolded bands in the photoemission spectra \cite{Marchenko2023}, however, not reflecting a factual moir{\'{e}}-dressed band structure. The direct spectroscopic evidence of the latter is thus the opening of gaps, which in turn is in conflict with the desired search for massless moiré-Dirac states.

%Moreover, given the ability to create Dirac states in artificial honeycomb-type moiré superlattices, it is reasonable to suppose that other emergent lattice symmetries can be formed, which may lead to even more robust quasiparticles with topological protection. This perspective, however, has (to the best of our knowledge) so far been overlooked.

\begin{figure*}[h!t!]
    \includegraphics{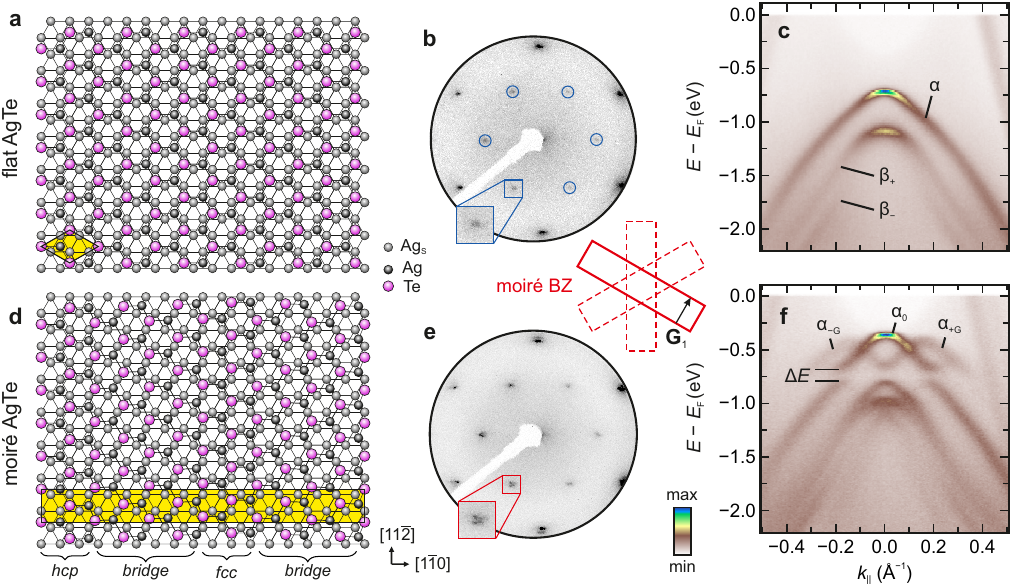}
    \caption{\textbf{Structural and electronic characterization of the flat (a-c) and moiré-modulated (d-f) honeycomb monolayer AgTe on Ag(111).} \textbf{a}, A ball model of the flat AgTe/Ag(111), where the surface unit cell is indicated by the yellow rhombus. \textbf{b}, The associated LEED image reveals the $(\sqrt{3}\times\sqrt{3})\mathrm{R}30^\circ$ -- AgTe superstructure. \textbf{c}, Surface electronic band structure of the flat AgTe -- as measured by ARPES -- features two hole-like dispersive bands dubbed $\alpha$ and $\beta_\pm$. \textbf{d}, A ball model of an exemplary $(17 \times \sqrt{3} )_\mathrm{rect}$ moiré AgTe superlattice on Ag(111), where the yellow rectangle highlights the enlarged surface unit cell. \textbf{e}, The corresponding LEED image reveals splitting of the diffraction spots, a characteristic of moiré superstructures. The moiré Brillouin zone is denoted by three red rectangles. \textbf{f}, Surface electronic band structure of the moiré AgTe as measured by ARPES. The observation of replica bands ($\alpha_{\pm \mathrm{G}}$) and gaps ($\Delta E$) evidences a substantial moiré potential. The data shown in \textbf{e},\textbf{f} approximately corresponds to a $(13 \times \sqrt{3} )_\mathrm{rect}$ superstructure, i.e., more compressed as the one depicted in \textbf{d}, where we choose a larger, less dense unit cell for illustration. The data corresponding to this structure is shown in Fig.~\ref{fig3}a. LEED and ARPES data were recorded at room temperature and $40\,\mathrm{K}$, respectively.}
    \label{fig1}
\end{figure*}

Here, we overcome these challenges and further exploit the potential of epitaxial moiré engineering for the creation, detection, and control of topological moiré-Dirac states.
We transform a 
%simple (and topologically trivial) surface system, an epitaxial honeycomb monolayer AgTe 
(topologically trivial) monolayer-substrate heterostructure (Fig.~\ref{fig1}a-c), into moiré-dressed topological Dirac matter.
Using ARPES, we find both sizable band gaps and Dirac-type band crossings in the emergent replica band structure.
%While the former proves the strength of the superlattice potential, which emerges from the spatially modulated monolayer-substrate interaction, the non-symmorphic symmetry of the long-range moiré lattice enforces and protects the latter.
%The former proves the strength of the superlattice potential and provides clear experimental evidence for a substantial decrease of the electrons kinetic energy. 
As being supported by comprehensive model considerations and symmetry analysis, we show that the crossing points belong to Dirac nodal lines, which are enforced and protected by the emergent non-symmorphic superlattice symmetry. The concomitant observation of gaps in turn proves the strength of the superlattice potential and particularly provides clear experimental evidence that the Dirac states are \textit{de facto} moiré-dressed.
Overall, our results allow us to identify the observed quasiparticle excitations as \textit{slow topological Dirac states}, which we directly demonstrate to be tunable, given their moiré-driven nature.
%Controllable tuning of the moiré superstructure allows us to directly manipulate the \textit{slow-Dirac dispersion} $E(\bold{k}) = \hbar v_g (\bold{k-k_0}) + E_0$, namely their energy $E_0$, group velocity $v_g$, and position in momentum space $\bold{k_0}$.

\section{moiré superlattice and replica band structure}

In Fig.~\ref{fig1}, we summarize the emergence of a line-moiré superlattice in the honeycomb monolayer AgTe on Ag(111). In the pristine form (Figs.~\ref{fig1}a-c), the low-energy electron diffraction (LEED) pattern  shows a $(\sqrt{3}\times\sqrt{3})\mathrm{R}30^\circ$ -- AgTe superstructure with sharp diffraction spots (Fig.~\ref{fig1}b). In this phase, the silver and tellurium atoms of the monolayer occupy the hexagonally closed-packed (\textit{hcp}) hollow sites of the underlying Ag(111) substrate. 
A typical ARPES measurement taken on this system is depicted in Fig.~\ref{fig1}c, showing two valence band features labeled as $\alpha$ and $\beta$. In both bands, the electronic states are mainly built from in-plane Te-$p_{x,y}$ orbitals with the key difference being the lateral orbital momentum-space alignment, referred to as in-plane orbital texture \cite{Uenzelmann2020}. The steep band crossing the Fermi level at $k_\parallel \approx \pm 0.4\,\mathring{\mathrm{A}}^{-1}$ corresponds to a Ag-$4s$ bulk band, backfolded by the superstructure, i.e., a pure photoemission final-state effect.

\begin{figure*}[h!t!]
    \includegraphics[scale=0.95]{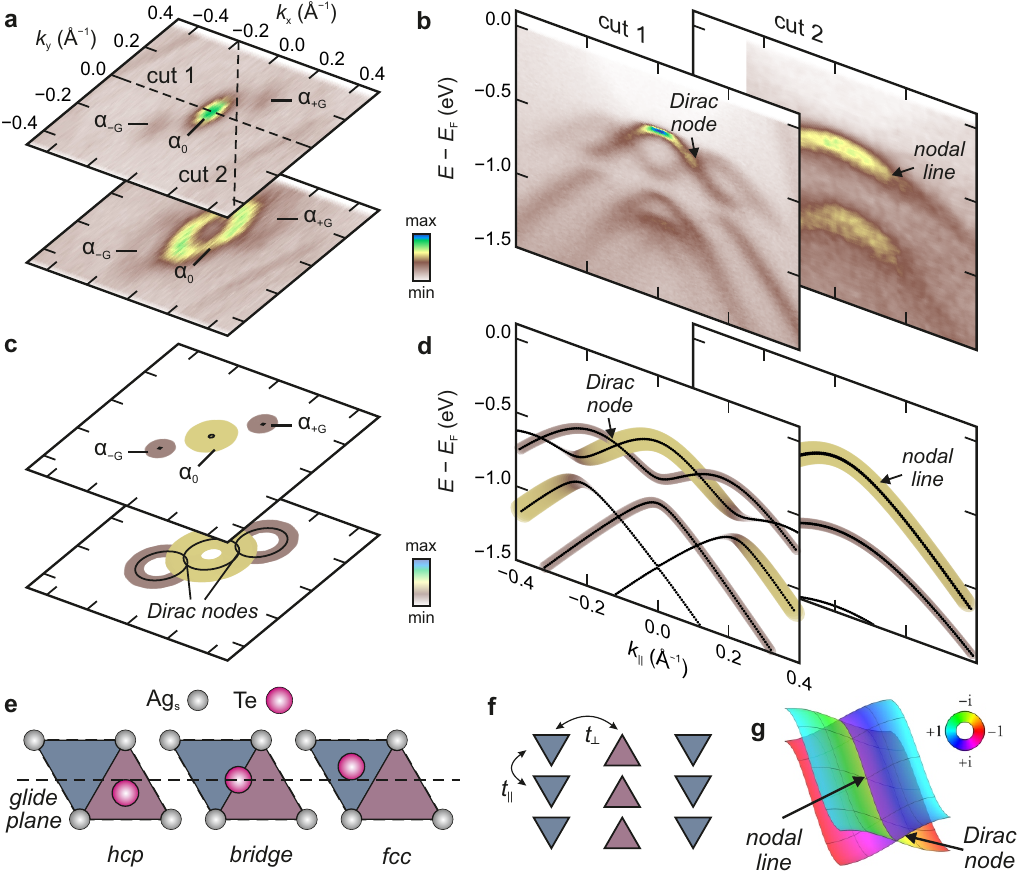}
    \caption{
    %\textbf{A thorough analysis of the electronic band structure of a moiré-modulated honeycomb monolayer AgTe on Ag(111) and the role of a glide plane. A closer look into the electronic band structure of a moiré-modulated honeycomb monolayer AgTe on Ag(111) and a symmetry analysis. Glide-plane enforced electronic structure in moiré-modulated honeycomb AgTe. Electronic band structure in moiré-modulated AgTe enforced by nonsymmorphic symmetry.} 
    \textbf{Formation of moir{\'{e}} band gaps and Dirac nodal lines from non-symmorphic moir{\'{e}} superlattice symmetry.}
    \textbf{a,b} ARPES data and \textbf{c,d} continuum model calculations (see text).
    \textbf{a}, \textbf{c}, Constant energy contours with the top and bottom slices correspond to energies of $-0.35\,\mathrm{eV}$ and $-0.46\,\mathrm{eV}$ relative to $E_\mathrm{F}$, respectively. \textbf{b}, \textbf{d}, Band structure cuts along the two distinct momentum paths indicated in \textbf{a} (cuts 1 and 2). Key electronic features, i.e., Dirac cone and Dirac nodal line are labeled across all plots for clarity. ARPES data was conducted at $40\,\mathrm{K}$.
    \textbf{e} Schematic illustrating the emergence of a glide-plane symmetry resulting from the formation of a moir{\'{e}} superlattice. For better visibility, we here restrict to the \textit{hcp}, \textit{bridge}, and \textit{fcc} sites only. %Small light-grey atoms correspond to Ag(111) subsurface atoms and the violet atom indicates the Te atom in uppermost layer.
    \textbf{f}, Schematic drawing of the coupled-chain tight-binding toy model for the uppermost two moir{\'{e}} bands. In analogy to \textbf{e}, the \textit{hcp} and \textit{fcc} sublattices are depicted as blue and violet triangles, respectively, and are related by non-symmorphic glide-mirror symmetry. The two kinetic hopping terms ($t_\parallel$) and ($t_\perp$) describe the intrachain and interchain coupling, respectively. \textbf{g}, Electronic tight-binding band structure derived from the model in \textbf{f}, giving rise to a topological Dirac nodal line, enforced and protected by non-symmorphic symmetry. Protection arises from the exchange of mirror eigenvalues, as being color-coded accordingly. 
    %\textcolor{red}{maybe indicate $k_x$ and $k_y$ in the BZ. Or is $k_x$ in (f) true at all? superlattice - 0.35 / 0.37 / 0.46 - add color code for ARPES}
    }
    \label{fig2}
\end{figure*}

\begin{figure*}[t!]
    \includegraphics{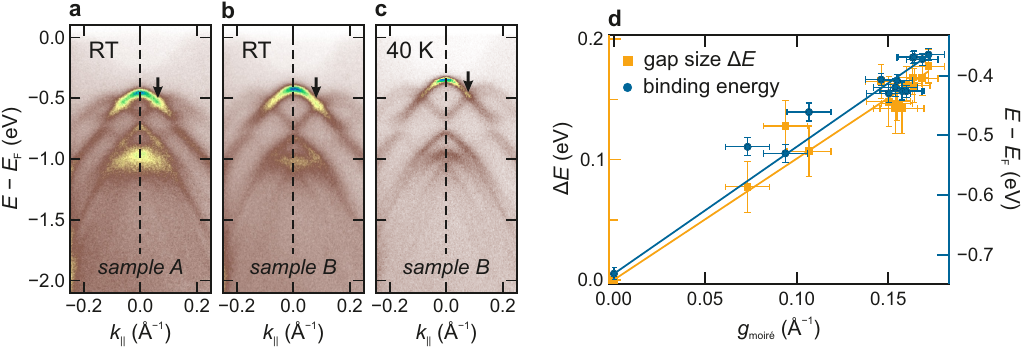}
    \caption{\textbf{Structural tuning of the moir{\'{e}} electronic band structure via the AgTe coverage and temperature.}
    %\textbf{a}, EDCs at $k_\parallel = 0\,\mathrm{\AA}^{-1}$ for three distinct \textit{moiré}-AgTe superlattices with increasing Te coverage (colored curves); horizontal dashes denote the emergence of moiré mini-gaps. An EDC for flat AgTe is included for reference (black curve).
    \textbf{a}--\textbf{c}, ARPES intensity maps recorded at room temperature (RT) (\textbf{a}, \textbf{b}) and at $40\,\mathrm{K}$ (\textbf{c}). Black arrows mark the symmetry-enforced Dirac crossings.
    %\textbf{e}, EDCs from the same \textit{moir{\'e}}-AgTe superlattice at two distinct temperatures (see labels and spectra \textbf{c}, \textbf{d}).
    \textbf{f}, Mutual scaling of the energy gap $\Delta E$ (yellow squares) and the energy of the valence band maximum (blue circles) against the moir{\'e} reciprocal lattice vector $g_\textrm{moir{\'{e}}}$. The linear trend highlights the systematic response of the AgTe band structure to the moir{\'{e}} superlattice periodicity.
    %\textbf{g} Variation of the moir\'e miniband ($\alpha_0$) bandwidth \textit{w} as a function of $g_\mathrm{moir\acute{e}}$. Error bars in \textbf{f} and \textbf{g} represent fitting uncertainties. \textcolor{red}{\textit{moir{\'e}}-AgTe superlattice - moiré periodicities - for numerous \textit{moir{\'e}}-AgTe periodicities}
    The error bars are derived from the confidence intervals by varying the center position (of the energy peak or Dirac point momentum) within reasonable limits. This is done at different points, determining the statistical uncertainty.}
    \label{fig3}
\end{figure*}

Increasing the coverage of the honeycomb layer slightly, the diffraction spots split into three sub-spots, as can be seen in Fig.~\ref{fig1}e.
The LEED pattern is characteristic of a long-range superlattice that emerges from a one-dimensional compression of the surface layer along the $[1\bar{1}0]$ (and symmetry-equivalent) directions  \cite{Soliton1987, Hammer2024}.
%That is, the reciprocal unit cell is one-dimensionally compressed along the $[1\bar{1}0]$ (and symmetry-equivalent) directions.
The hexagonal symmetry of the substrate results in three rotational domains, each rotated by $120^\circ$ relative to the others.
A real space structure model of a single domain is depicted in Fig.~\ref{fig1}d, which agrees with the atomic structure relaxed in our density functional theory (DFT) calculations (see Supplementary Information). Through the compression of the monolayer, not only the \textit{hcp} hollow sites but also the face-centered-cubic (\textit{fcc}) hollow sites become occupied. The \textit{hcp}-stacked and \textit{fcc}-stacked areas, in turn, are smoothly separated by soliton-type domain boundaries \cite{Soliton1987} in which, particularly, the bridge sites of the underlying Ag-layer are covered. The appearance of different stacking areas is a typical characteristic of moiré superlattices, in graphene and related 2D heterostructures known as AA and AB sublattice stackings. In analogy, one can consider the 1D structure at hand as a line-moiré-type superlattice \cite{Hammer2024} that emerges from the compression of the AgTe layer with respect to the Ag(111) substrate.
\\
An ARPES spectrum recorded from a moiré sample is shown in Fig.~\ref{fig1}f. The overall shape of the parent band structure (Fig.~\ref{fig1}c) can still be recognized, indicating a close structural relation of the two phases on a short-range atomic scale. However, there are distinctive differences: First, the bands are shifted by $\approx 400\,\mathrm{meV}$ towards the Fermi level. We interpret this as a reduction of the initial charge transfer from the substrate into the AgTe monolayer, leading to a complete occupation of the Te-derived bands from the outset. The physical origin of this effect will play a minor role in the following, but we later get back to the energy shift, acting as a useful indicator to distinguish different moiré phases (see Fig.~\ref{fig3}). 
%in the uncompressed AgTe monolayer. 
More relevant here is the appearance of replica bands $\alpha_{\pm \mathrm{G}}$ next to the original main band, from now labeled as $\alpha_0$ (Fig.~\ref{fig1}f and Fig.~\ref{fig2}a,b).
These mini-bands emerge from the line-moiré superlattice and are shifted by the first reciprocal lattice vectors $\bold{G_1}$ of the moiré mini-BZ. 
%(compare sketch in Fig.~\ref{fig1}).
Note that the reason the band maximum of the replica bands is at slightly lower energy as compared to the main band is that energy-momentum cuts are not taken directly along the high-symmetry direction of the moiré BZ (see Supplementary Information).
We furthermore note that replica bands and gaps $\Delta E$ are only found in the $\alpha$ band, whereas the $\beta$ states seem to be unaffected by the moiré lattice. As mentioned before, the two bands differ mainly by the in-plane orbital texture \cite{Uenzelmann2020}, and we therefore conclude that the lateral orbital alignment influences the moiré potential strength of the respective states. The potential is determined by the spatially varying local interlayer coupling $V(\bold{r})$ between the AgTe monolayer and the subsurface silver layer, which, in general, is altered by the involved orbitals.

The observation of sizable band gaps of about $\Delta E \approx 160\,\mathrm{meV}$ (specific to this particular moiré phase, cf. Fig.~\ref{fig3}) is a direct measure of the strength of the superlattice potential. Accordingly, the kinetic energy of the electrons --- propagating in the direction in which the AgTe layer is compressed --- is substantially quenched with a bandwidth of $W \approx 200\,\mathrm{meV}$, which is one order of magnitude less than for the flat AgTe layer. The moiré-dressed electronic states can thus be considered as highly one-dimensional.
Next to the band gaps, at $k_\parallel \approx \pm 0.1\,\mathring{\mathrm{A}}^{-1}$ we observe clear Dirac-like features with linear crossings between the main band $\alpha_0$ and the next replica bands $\alpha_{\pm \mathrm{G}}$ (Figs.~\ref{fig1}f and \ref{fig2}b).
Importantly, crossing points in the replica bands observed in ARPES often turn out as trivial photoemission final state backfoldings, because band gaps only allow for a clear identification as initial state replicas \cite{Marchenko2023}.
However, in our case, we clearly find the same bands that form the Dirac crossings, $\alpha_0$ and $\alpha_{\pm \mathrm{G}}$, are themselves involved in the formation hybridization gaps at other momenta. This allows us to identify also the Dirac crossings as \textit{de facto} property of the initial state band structure.

%Importantly, the entire observed replica band structure can therefore be identified as an \textit{de facto} initial state effect and does not emerge from trivial photoemission final state backfolding. This may likewise give rise to sidebands in the ARPES spectra, which, however, would not show any band gaps.

Overall, our ARPES data unveil the electronic structure to be decisively altered by the 1D moiré superlattice.
Across the latter, the electronic states can be described as \textit{slow}, in terms of their superlattice-driven reduced kinetic energy, but \textit{massless}, in that they form emergent Dirac crossings. This is the key experimental observation of this study. Strongly resonating with recent works on moiré-driven \textit{slow Dirac fermions} \cite{Yang2025, Ma2025, Herbut2006, Biedermann2025, Biedermann2026}, we here provide the first momentum-resolved evidence of this paradigm.
In the following, we will substantiate our experimental findings with model considerations and a profound symmetry analysis. Finally, we will directly demonstrate a key advantage of the moiré-based nature of the observed states, namely their controlled manipulation via the superlattice as a tuning knob.

%Finally, we point out that the almost completely vanishing spectral weight within the gaps clearly indicates that the photoemission signal does not be an average of all three rotational moiré domains, which would inevitably produce in-gap spectral weight 
%
%Although one expects photoemission to spatially average over the three rotational domains, we find the ARPES signal to be predominantly sensitive to a single domain. This is likely caused by photoemission matrix element effects, seemingly suppressing the signal from the other rotational domains. In particular, we find clear moiré hybridization gaps $\Delta E$ between the $\alpha_{\pm \mathrm{G_1}}$ bands and between $\alpha_{\pm \mathrm{G_2}}$ and $\alpha_0$ with an almost completely vanishing spectral weight in the gaps. These would be merely visible in an entirely domain-averaged signal (see Supplementary Information).
%This appears from the constant energy contour in Fig.~2xx, showing only two instead of six replica bands. Moreover,

\section{Formation of band gaps and Dirac nodal lines}

To elaborate on the microscopic mechanisms underlying the formation of the superlattice potential, band gaps, and Dirac nodes, we devise a continuum model to describe the moiré band structure (Fig.~\ref{fig2}). 
%Instead of a tight-binding-based continuum model, widely used for the description of the mini-bands in twisted bilayer graphene or transition metal dichalcogenides, 
We here use an modified nearly-free-electron model with plane-wave basis wave function $\psi_\bold{G_n}(\bold{r}) \propto \mathrm{e}^{i (\bold{k+G_n}) \bold{r} }$. Given the vanishing influence on the $\beta$ band, we restrict the model to $\alpha$. The energies of the pristine bands $E_\alpha (\bold{k_\parallel + G_n})$ (i.e., without the influence of the potential) are phenomenologically determined by the band dispersion of $\alpha$ around the $\overline{\Gamma}$ point (see Supplementary Information). We also neglect the slight hexagonal warping of the band and treat the pristine AgTe monolayer as isotropic (compare Figs.~\ref{fig2}a,c). Overall, our model does not aim for a quantitatively perfect match with the measured band dispersion but rather to address the relevant physical mechanisms and to capture the apparent hot spots in the moiré band structure. 
The superlattice potential is written as a Fourier series 
$V(\bold{r}) = \sum_n V(\bold{G_n}) \mathrm{e}^{i \bold{G_n} \bold{r} }$,
which effectively describes the spatially varying interlayer coupling by the Fourier coefficients $V(\bold{G_n})$, with the $n$-th reciprocal lattice vectors $\bold{G_n}$. Moreover, to capture the ARPES data, the model eigenstates are projected onto plane waves --- reflected by the color and dot sizes in the model calculations --- which yields the ARPES spectral weights \cite{Voit2000, Lisi2021, Gatti2023} (see also Supplementary Information).

%reflect the local interlayer hybridization terms $\omega_\mathrm{hcp,fcc}$ and $\omega_\mathrm{b}$ at the fcc, hcp, and bridge stacking configurations, respectively (see Fig.~2xx). In particular, we consider the first two terms with
%$|V_1| = | \omega_\mathrm{hcp} -  \omega_\mathrm{fcc}|$
%and
%$|V_2| = | \omega_\mathrm{hcp,fcc} -  \omega_\mathrm{b}|$ \textcolor{red}{[footnote; sign of the potential is unknown].}

In Fig.~\ref{fig2}a-d, we compare our experimental data with the model band structure.
Despite the simple character of the continuum model, we get a very good agreement comparing both constant energy contours in Figs.~\ref{fig2}a,c and $E$-$k$ band structure cuts in Figs.~\ref{fig2}b,d. Further comprehensive comparison of experimental data and model calculations --- including various momentum-directions and consideration of rotational domains --- is provided in the Supplementary Information.
With the used model parameters, on which we will further elaborate in the following, we can reproduce all relevant aforementioned features. In particular, we find sizable band gaps and Dirac nodes, highlighted accordingly in 'cut$\,1$' in Figs.~\ref{fig2}b,d.
The gaps appear as avoided crossings among the $\alpha_{\pm \mathrm{G}}$ bands themselves as well as between $\alpha_0$ and the next replica bands $\alpha_{\pm 2 \mathrm{G}}$.
Note, that the latter have an almost completely vanishing photoemission spectral weight and are visible only in the close vicinity of the avoided crossing point where the eigenstates are sufficiently mixed with $\alpha_0$ upon hybridization.
The Dirac nodes can be found at the mini Brillouin zone (mBZ) boundary where the $\alpha_0$ main band and the first replicas $\alpha_{\pm \mathrm{G}}$ cross.
Interestingly, the crossings emerge not at singular points in momentum space, but rather as a line degeneracy, i.e., Dirac nodal lines, pinned to the mBZ boundary. This can be seen from the corresponding 'cut$\,2$' along the mBZ edge in Figs.~\ref{fig2}b,d, where the bands (that touch at single points across, e.g., 'cut~1') are completely degenerate.

To reproduce the experimentally observed moiré-induced Dirac crossings and band gaps, we find two relevant sets of model parameters: First of all, the odd Fourier coefficients have to be set to zero because finite values of $V(\bold{G_\mathrm{odd}})$ would immediately lead to a gap opening at the zone boundaries, contradicting the experimental observation.
Furthermore, the second Fourier component accounts for the observed gap opening, i.e., we derive $|V(\bold{G_2})| = \Delta E/2 \approx 80\,\mathrm{meV}$ from the experimental data.
Higher-order even Fourier components are not strictly prohibited, but we find their magnitudes to be at least below the detection limit, and we set $V(\bold{G_{|n|>2}})=0$.
On that basis, we can reconstruct the full effective moiré potential, which has the periodicity of half a superlattice period. To understand this from a microscopic perspective, we take a closer look at the local adsorption site stacking area, i.e., Te atoms sitting at the \textit{hcp}-, \textit{fcc}-hollow, and \textit{bridge} sites, as depicted in Fig.~\ref{fig2}e. The second Fourier component yields the local potential difference between the hollow and bridge site coupling terms $V(\bold{r_{hcp,fcc}}) - V(\bold{r_{bridge}}) = V(\bold{G_2})$. It becomes immediately evident that the local environments at these two kinds of sites are considerably different, which gives rise to a finite potential difference of $|V(\bold{G_2})| \approx 80\,\mathrm{meV}$ (the value being specific to this particular moiré phase). This causes the electrons to be trapped at or between the \textit{bridge} site areas, depending on the sign of the potential, which is not directly accessible from the experimental band dispersion. The trapping, in turn, causes the quenching of kinetic energy, a reduced band width, and thus effectively one-dimensional character of the electronic states.
One can consider the line-moiré structure effectively as a coupled-chain system of alternating \textit{hcp}- and \textit{fcc}-type chains (Fig.~\ref{fig2}f) with a weak coupling ($t_\perp$) that is determined by the moiré potential.
\\
%, with the intrachain hopping $t_\parallel$ being given by the short-range atomic orbital overlaps and the interchain coupling
Comparing the \textit{hcp}- and \textit{fcc} sites with each other, one finds the same local environments if the bilayer system is considered of AgTe and a single subsurface Ag(111) layer. In this case, the interlayer coupling is $V(\bold{r_{hcp}}) = V(\bold{r_{fcc}})$, yielding
\begin{align}
\label{Eq1}
V(\bold{r_{hcp}}) - V(\bold{r_{fcc}}) = 2 \sum V(\bold{G_\mathrm{odd}})=0 . 
\end{align}
That is, there is an inherent connection between the local adsorption-site symmetries in the moiré superlattice and the emergence of Dirac nodal lines.

\section{Band topology from non-symmorphic superlattice symmetry}

We will now show that these phenomenological observations, such as the occurrence of Dirac nodal lines, in particular, are related to strict symmetry constrains, enforcing and protecting the observed line nodes.
Key ingredient is the non-symmorphic glide mirror symmetry $\tilde{M}_y$ of the combined heterosystem, i.e., the compressed AgTe surface layer and next Ag(111) subsurface layer below (see Figs.~\ref{fig1}d and \ref{fig2}e). The glide mirror symmetry comprises a half moiré-lattice translation $t_x/2$ and a mirror operation $M_y$.
It has been shown in various earlier works that such symmetries give rise to topological Dirac semimetal phases \cite{Young2015, Schoop2016, Wilde2021, Hirschmann2021}. In fact, this is the case for the heterostructure at hand, for which our symmetry analysis yields a Dirac nodal line. That is, $\tilde{M}_y$ in combination with time-reversal symmetry $\mathcal{T}$ square to 
$(\tilde{M}_y \mathcal{T})^2 = \tilde{M}_y^2 \mathcal{T}^2 = \mathrm{e}^{i q_x a_\textrm{moir{\'{e}}}}
$. Here, $q_x$ refers to the wave vector along the compressed $[1\bar{1}0]$ direction in the reference frame of a single rotational domain.
The anti-unitary symmetry $\tilde{M}_y \mathcal{T}$ thus fulfills Kramers' theorem at $q_x = \pi/a_\textrm{moir{\'{e}}}$ and enforces a band degeneracy along the invariant $k$-points which are $(q_x,q_y) = (\pi/a_\textrm{moir{\'{e}}},q_y)$, i.e, the entire edge of the rectangular mBZ is degenerate.
The topological character of this robust protection is encoded in the exchange of mirror eigenvalues across the nodal line, which are winding through the mBZ, as depicted in Fig.~\ref{fig2}g.

The preliminary phenomenological assumptions that were based on the comparison of the continuum model with the experimental data in the previous section are thus in fact deeply supported by symmetry arguments. In other words, symmetry indeed requires $V(\bold{G_\mathrm{odd}})=0$, in order to protect the Dirac nodal line (Fig~\ref{fig2}b,d).
This can be related to the preceding symmetry argument: Glide mirror symmetry maps the \textit{hcp} sites to the \textit{fcc} sites, which for the local interlayer coupling terms yields $V(\bold{r_{hcp}})=V(\bold{r_{fcc}}) \Leftrightarrow V(\bold{G_\mathrm{odd}})=0$.
This condition can be derived analytically when introducing glide mirror symmetry to the moiré potential 
\begin{equation}
\label{Eq_Sym}
V(\tilde{M}_y\bold{r}) = V(\bold{r}) \Rightarrow V(\bold{G_\mathrm{odd}})=0~,
\end{equation}
which is set out in more detail in the Supplementary Information. 
Our analysis thus combines aspects of both the global superlattice symmetry $\tilde{M}_y$ and the periodic long-range moiré potential $V(\bold{r})$. The latter, in turn, is built from the local interlayer couplings at different stacking areas (see Eq.~\ref{Eq1} and Fig.~\ref{fig2}e).
\\
Taken together, the observed Dirac nodal lines are topologically protected by the non-symmorphic symmetry, emergent in the moiré superlattice and absent in the parent (non-moiré) AgTe monolayer.
The observation of symmetry-enforced nodal features alone is not new but has been reported in various experimental studies, such as, in Refs.~\cite{Schoop2016, Wilde2021, Figgemeier2025}.
The key advantage in our work is that we highlight their feasibility on a moiré superlattice, demonstrating that the exploration of novel topological Dirac states does not require the search for (and prediction of) novel quantum material compounds and their single-crystal synthesis. In contrast, the relevant lattice symmetries  \cite{Bradlyn2017} can emerge in artificial electronic quantum materials, i.e., particularly moiré systems, as elaborated here.

%Given that we have a 2d systen, this nodal line can be considered as a 1d version of a nodal plane, which can be found in 3d topological semimetals \cite{Wilde2021}.

\section{Moiré-controlled tuning of the Dirac states}

%\vspace{0.5cm}
%\textbf{\large Moiré-controlled tuning of the Dirac states}
%\\
Finally, we focus on the naturally arising question of whether the band structure can be modified through the moiré structure. In analogy to the twist-angle tuning in twisted multilayer vdW heterostructures, the tuning knob at hand is the epitaxial modification of the coverage and thereby compression of the AgTe monolayer.
To this end, we realized the preparation of samples with various coverages and thus different superlattice periodicity. ARPES measurements of two exemplary samples labeled \textit{sample~A} and \textit{sample~B}, both taken at room temperature (RT), are shown in Fig.~\ref{fig3}a,b. \textit{Sample~B} has a higher coverage, which leads to an increased compression and thus smaller real space moir{\'{e}} wavelength. 
Consequently, the Dirac points in \textit{sample~B} are at higher wave vectors ($\pm0.082 \pm 0.012 \,\mathring{\mathrm{A}}^{-1}$) as compared to \textit{sample A} ($\pm0.062 \pm 0.012 \,\mathring{\mathrm{A}}^{-1}$). Moreover, we systematically find the bands to be closer at the Fermi level for larger mBZs, i.e., increased compression. As described earlier, this energy shift serves as a useful additional indicator for different moiré phases.
\\
Next to the coverage, we have examined the influence of the sample temperature on the electronic band structure. In Fig.~\ref{fig3}c, we show an ARPES measurement taken on the same sample as in Fig.~\ref{fig3}b, but measured at lower temperature of $T=40\,$K. In spite of the same nominal coverage, we find the mBZ enlarges, the Dirac points are at slightly higher parallel momenta ($\pm0.088 \pm 0.010 \,\mathring{\mathrm{A}}^{-1}$), and the bands are rigidly shifted towards $E_\mathrm{F}$, all of which directly indicates a temperature-driven change of the moir{\'{e}} phase. Moreover, as being most clearly seen by comparing these two measurements, also the gap size increases at lower temperature, proving the moir{\'{e}} potential to scale with superlattice periodicity. Notably, we find this temperature-dependent phase transition to be completely reversible, showing that the effect is not caused by, e.g., a thermal-energy-induced Te desorption or sublimation, which would also not be expected at temperatures as low as RT. We note that the mechanism underlying the observed temperature effect is not yet fully understood. The nature of the formation of the line-moir{\'{e}} structure is quite complex. It is governed by the delicate interplay between the lateral interatomic forces within the AgTe layer and the interaction with the substrate, to both of which temperature may also contribute. We here refrain from further speculations and focus on the phenomenology, i.e., temperature appears as an additional tuning knob of the moir{\'{e}} system. 
\\
In Fig.~\ref{fig3}d, we present the condensed results observed from various prepared moir{\'{e}} phases. The plot shows the dependence of the gap size $\Delta E$ and energy of the valence band maximum relative to $E_\mathrm{F}$ as a function of the size of the reciprocal moiré lattice vector $g_\textrm{moir{\'{e}}}=\frac{2\pi}{a_\textrm{moir{\'{e}}}} = |\textrm{G}_1|$. Here, we consider the pristine (non-moir{\'{e}}) AgTe phase as the limit of an \enquote{infinite moir{\'{e}}} wavelength $a_\textrm{moir{\'{e}}}$, which yields $g_\textrm{moir{\'{e}}}=0$.
Phenomenologically, we find a kind of universal scaling behavior, with both the gap size and the energy shift being almost perfectly proportional to the size of the mBZ. The solid lines correspond to the linear fits to the respective data sets.
In general, our data shows that the strength of the moiré potential as well as the Dirac point momentum --- reflected by $\Delta E$ and $g_\textrm{moir{\'{e}}}$, respectively --- can be almost continuously tuned by a factor of $\approx 2$. Likewise, other band parameters, such as the bandwidth
%\textcolor{red}{real space extension of the moiré Wannier orbitals}, 
are controlled by the superlattice (see Supplementary Information), altogether underpinning the high degree of tunability in the system.

\section{Discussion and Outlook}

In sum, our data provide spectroscopic evidence for the formation of robust, gapless, and tunable Dirac nodal lines in a moiré-like superlattice. While direct momentum-resolved evidence of moiré-Dirac states in twisted vdW layers is still outstanding, the concomitant observation of mini-gaps and band crossings shown here serves as proof of principle from a broader perspective. The presented epitaxial heterostructure appears as an ideal model platform for this purpose, exhibiting long-range order and non-symmorphic symmetry.
Beyond its role as a model system, the moiré-AgTe/Ag(111) system has particularly interesting aspects on its own.
First, our theoretical analysis shows that the system’s topology is even richer: Spin-orbit coupling transforms the nodal lines into Dirac hourglass states that may be resolvable in spin-resolved ARPES measurements.
%First, we find (in theory) that the topology of the system is even richer, i.e., spin-orbit coupling lifts spin degeneracy at the nodal lines transforming them into gapless Dirac hourglass states, which may be resolvable in spin-resolved ARPES measurements.
Those that have already been carried out on pristine AgTe/Ag(111) confirm a small spin-splitting of the $\alpha$-band, not detectable in spin-integrated measurements \cite{Geldiyev2026}. Additional to the Dirac states, topology also manifests itself at the observed mini-gaps in the form of quantum metric, and the experimental investigation of this is currently being put into practice \cite{Uenzelmann2026}.
Finally, given the fact that the presented moiré system is tunable by sample temperature, we suppose that the structure may likewise be triggered by strong, ultrashort light pulses, such that the topological moiré band structure could be controlled and imaged on ultrafast timescales.

\bibliographystyle{apsrev4-2}
\bibliography{Moire}

%\vspace{5cm}

\begin{section}{Acknowledgements}
R.G., M.P.T.M., and M.Ü. thank A. Crepaldi for valuable inputs.
M.M.H. thanks A. Furusaki for helpful discussions.
The Würzburg authors thank the
Deutsche Forschungsgemeinschaft (DFG) supporting this through the Würzburg-Dresden Cluster of Excellence \textit{ctd.qmat} (EXC 2147, Project ID 390858490), SFB1170 'ToCoTronics', and Project RE1469-13-2 at JMU Würzburg.
M.M.H. was supported by the RIKEN Special Postdoctoral Researcher Program.
A.C. acknowledges support from PNRR MUR project PE0000023-NQSTI. A.C. further acknowledges the Gauss Centre for Supercomputing e.V. (https://www.gauss-centre.eu) for funding this project by providing computing time on the GCS Supercomputer SuperMUC-NG at Leibniz Supercomputing Centre (https://www.lrz.de).
J.S. acknowledges funding from the European Union’s Horizon Europe research and innovation programme under the Marie Skłodowska-Curie grant agreement No 101209345-ART.QM funded by the European Union. Views and opinions expressed are, however, those of the author(s) only and do not necessarily reflect those of the European Union or European Research Executive Agency. Neither the European Union nor the granting authority can be held responsible for them.
\end{section}

%\vspace{1cm}

\begin{section}{Author contributions}
    R.G. and M.P.T.M. prepared the samples, carried out the ARPES measurements, and analyzed the experimental data with support from B.G. and M.Ü..
    The continuum model has been set up by B.G. and M.Ü. with input from F.R., R.G., and M.P.T.M.. J.S. and M.Ü. have analyzed the spectral weights. A.C. and D.D.S. did the DFT calculations.
    M.M.H. has performed the symmetry analysis and tight-binding modeling and provided direct inputs to the writing of the manuscript.
    M.Ü. was responsible for the project coordination and wrote the manuscript with input from all co-authors. The overall responsibility is shared by M.Ü. and F.R..
\end{section}

\iffalse

Further non-sym.:
1D directional massless Dirac vdW long-range order and CDW induced nodal lines
\cite{Yang2020, Sarkar2023} \\
%
1D Moiré
\cite{Kawakami2026} \\
%
Moiré Theory
\cite{Mele2010, Suarez2010, Bistritzer2011, Wu2018, Wu2019, Balents2020}

\fi

\end{document}